\documentclass[twocolumn,english]{article}
\usepackage[T1]{fontenc}
\usepackage[latin9]{inputenc}
\usepackage{amsthm}
\usepackage{amsmath}
\usepackage{graphicx}

\makeatletter
\numberwithin{equation}{section}
\numberwithin{figure}{section}
\newcommand{\code}[1]{\texttt{#1}}

\@ifundefined{showcaptionsetup}{}{%
 \PassOptionsToPackage{caption=false}{subfig}}
\usepackage{subfig}
\makeatother

\usepackage{babel}
\begin{document}

\title{Towards a Java Subtyping Operad\footnote{This paper has been accepted for publication at FTfJP'17~\cite{AbdelGawad2017a}.}}

\author{Moez A. AbdelGawad\\Informatics Research Institute, SRTA-City, Alexandria, Egypt\\\code{moez@cs.rice.edu}}

\maketitle

\begin{abstract}
	The subtyping relation in Java exhibits self-similarity. The self-similarity
	in Java subtyping is interesting and intricate due to the existence
	of wildcard types and, accordingly, the existence of three subtyping
	rules for generic types: covariant subtyping, contravariant subtyping
	and invariant subtyping. Supporting bounded type variables also adds
	to the complexity of the subtyping relation in Java and in other generic
	nominally-typed OO languages such as C\# and Scala.
	
	In this paper we explore defining an operad to model the construction
	of the subtyping relation in Java and in similar generic nominally-typed
	OO programming languages. Operads, from category theory, are frequently
	used to model self-similar phenomena. The Java subtyping operad, we
	hope, will shed more light on understanding the type systems of generic
	nominally-typed OO languages.
\end{abstract}

\section{\label{sec:Introduction}Introduction}

\def\JSO{\mathcal{JSO}}
\def\JSM{JSM}
The addition of generics to Java~\cite{JLS05,JLS14} made the subtying
relation in Java more intricate, particularly after adding wildcard
types~\cite{Torgersen2004} that can be passed as type parameters
to generic types.

Wildcard types in Java express so-called \emph{usage-site} variance
annotations\emph{. }Due to supporting wildcard types, the subtyping
relation between generic types in Java is governed by three rules:
covariant subtyping, contravariant subtyping, and invariant subtyping.
Covariant subtyping causes different generic types parameterized with
types to be in the \emph{same} \emph{subtyping} \emph{relation} with
respect to each other as their type parameters are, using the wildcard
parameter `\code{?~extends~Type}'. Contravariant subtyping causes
different generic types parameterized with types to be in the \emph{opposite}
\emph{subtyping} \emph{relation} with respect to each other as their
type parameters are (\emph{i.e.}, generic types parameterized by \emph{subtype}
type parameters become \emph{supertypes}), using the wildcard parameter
`\code{?~super~Type}'. Invariant subtyping causes different generic
types parameterized with types to be in \emph{no} \emph{subtyping}
\emph{relation} with respect to each other, regardless of the subtyping
relation between their type parameters, using the type parameter `\code{Type}'
(\emph{i.e.}, in Java, invariant subtyping is the default subtyping
rule between generic types when there is no `\code{?~extends}'
and `\code{?~super}' annotations).

Further adding to the intricacy of the subtyping relation is the fact
that type variables of a generic class\footnote{In this paper Java interfaces are treated as similar to abstract classes.}
can have upper bounds, restricting the set of types that can be passed
as type parameters in instantiations of the generic class. We explore
in more detail the implications of the generic subtyping rules on
the subtyping relation in Java in Section~\ref{sub:Subtyping-in-Java}.

The subtyping relation in other industrial-strength generic nominally-typed
OOP languages such as C\#~\cite{CSharp2015} and Scala~\cite{Odersky14}
exhibits similar intricacy. C\# and Scala support another kind of
variance annotations (called \emph{declaration-site} variance annotations)
as part of their support of generic OOP. Both kinds---\emph{i.e.},
usage-site and declaration-site variance annotations---have the same
self-similarity effect on the generic subtyping relation in nominally-typed
OOP languages.

As we describe in Section~\ref{sec:Related-Work} in more detail,
the introduction of wildcard types (and variance annotations, more
generally) in mainstream OOP, even though motivated by earlier research,
has generated much additional interest in researching generics and
in having a good understanding of variance annotations in particular.
In this paper we augment this research by presenting an operad for
modeling the subtyping relation in Java, exhibiting and making explicit
in our operad the self-similarity in the definition and construction
of the relation, with the expectation that our operad will apply equally
well to subtyping in other OO languages such as C\# and Scala.

This paper is structured as follows. In Section~\ref{sec:Background}
we present a demonstration of the intricacy and self-similarity of
the subtyping relation in Java, followed by a brief introduction to
operads. Then, in Section~\ref{sec:JSO}, we present $\JSO$, our
operad for modeling the subtyping relation in Java and other generic
nominally-typed OOP languages. In Section~\ref{sec:Applying-JSO}
and Appendix~\ref{sec:Demonstration-of-JSO} we present examples
of the application of $\JSO$ that demonstrate how it works. In Section~\ref{sec:Related-Work}
we discuss some research that is related to ours. We conclude in Section~\ref{sec:Conclusion-and-FW}
by discussing some conclusions we made and discussing some research
that can be built on top of this work.

\section{\label{sec:Background}Background}

In this section we give a simple example of how the subtyping relation
in a Java program can be constructed iteratively, based on the type
and subtype declarations in the program. We follow that by a brief
introduction to operads.

\subsection{\label{sub:Subtyping-in-Java}Subtyping in Java}

To explore the intricacy of the subtyping relation in Java, let's
consider a simple example---probably the most basic example---of a
generic class declaration.

Assuming we have no classes or types declared other than class \code{Object}
(whose name we later abbreviate to \code{O}), with a corresponding
type that has the same name, then the generic class declaration

\code{class C<T> extends Object \{\}}\\
results in a \emph{subclassing} relation in which class \code{C}
is a subclass of \code{O}.

For subtyping purposes in Java, it is useful to also assume the existence
of a special class \code{Null} (whose name we later abbreviate to
\code{N}) that is a subclass of all classes in the program, and whose
corresponding type (that is sometimes called \code{NullType}~\cite{JLS05,JLS14}\footnote{As of Java 8.0, the \code{Null} type is inexpressible in Java. It
is needed, however, during type checking, particularly when checking
the types of some expressions that involve polymorphic method type
inference and wildcard types~\cite{JLS05,JLS14}.}, but which we also call the type \code{Null} for consistency), hence,
is a subtype of all Java object types (\emph{a.k.a.}, reference types).
Accordingly, the full subclassing relation (based on the earlier declaration
of class \code{C}) looks as in Figure~\ref{fig:Subclassing}.

\noindent \begin{center}
\begin{figure}
\noindent \begin{centering}
\subfloat[\label{fig:Subclassing}Subclassing]{\protect\includegraphics[scale=0.6]{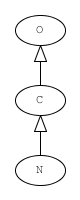}

}~~~~~\subfloat[\label{fig:Initial-Subtyping}Subtyping between rank 0 types]{\protect\includegraphics[scale=0.6]{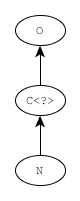}

}
\par\end{centering}

\protect\caption{}
\end{figure}

\par\end{center}

We now informally describe how the generic subtyping relation in our
Java program can be constructed iteratively, based on the mentioned
assumptions and the declaration of generic class \code{C}. For simplicity,
we further assume that a generic class takes only one type parameter,
and that type variables of all generic classes have type \code{O}
(\emph{i.e.}, \code{Object}) as their (upper) bound.

Given that we have at least one generic class, namely \code{C}, to
construct the generic subtyping relation we should note first that
the relation will have an \emph{infinite} number of types, since generic
types can be arbitrarily nested. As such, to construct the infinite
subtyping relation we go in iterations, where we start with a finite
first approximation to the relation then, after each iteration, we
get a step closer to the full relation.

The input to the first iteration of the construction process will
be the subtyping relation between a finite set of initial types (\emph{i.e.},
ones having rank 0) that are all defined directly using the subclassing
relation, then the new types that get constructed in the first iteration
are of rank 1, and they get fed as part of the input subtyping relation
to the second iteration of the process, and so on.

To demonstrate the workings of this iterative process in more detail,
we use the class \code{C} declaration above and assume, momentarily,
the existence of the covariant subtyping rule only (\emph{i.e.}, allow
only type parameters of the form `\code{?~extends~Type}'). Later
on we show how types constructed for the two other generic subtyping
rules (contravariant, using `\code{?~super~Type}' type parameters,
and invariant, using `\code{Type}' type parameters) and the resulting
subtyping relations get incorporated in the construction process.

Types of rank 0---the input to the first iteration of the construction
process---are defined immediately, using the subclassing relation,
where each non-generic class has a type corresponding to the non-generic
class with the same name as the class, while every generic class (only
class \code{C} in our example) gets passed the default `\code{?}'
type parameter to define its corresponding rank 0 type. The subtyping
relation between rank 0 types will always be exactly \emph{the} \emph{same}
as the subclassing relation between the classes used to define the
types. As such, the diagram for the initial subtyping relation will
look very similar to that of the subclassing relation, except that
it is for the subtyping relation between object types rather than
the subclassing relation between classes. (See Figure~\ref{fig:Initial-Subtyping}.)

To construct the second, more accurate version of the full subtyping
relation, we use the generic classes and the available subtyping rules
(only covariant subtyping, for the moment) to construct new types.
In relation to each other, these new types (of rank 1) will be in
the new version of the subtyping relation based on the available subtyping
rule(s).

More concretely, using the generic class \code{C} and the covariant
subtyping rule, in the first iteration of the construction process
the rank 1 types

\begin{figure}
\noindent \begin{centering}
\subfloat[\label{fig:Covariant-Subtyping}Subtyping (between types of rank 0
and 1) due to \emph{covariant} subtyping]{\protect\includegraphics[scale=0.6]{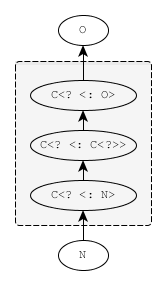}

}~~~\subfloat[\label{fig:Contravariant-Subtyping}Subtyping (between types of rank
0 and 1) due to \emph{contravariant} subtyping]{\protect\includegraphics[scale=0.6]{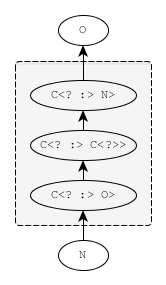}

}
\par\end{centering}

\protect\caption{}
\end{figure}

\code{C<?~<:~O>}~~~~~and~~~~~\code{C<?~<:~C<?>\textcompwordmark{}>}~~~~~and~~~~~\code{C<?~<:~N>}\\
(where we abbreviate \code{extends} to `\code{<:}'; we later also
abbreviate \code{super} to `\code{:>}') are constructed, and these
three types will have the \emph{same} subtyping relation as that between
their non-annotated type parameters (namely, \code{O}, \code{C<?>},
and \code{N}, respectively) in the first version of the subtyping
relation. As such, the subtyping relation resulting from the first
iteration of the construction process will be as in Figure~\ref{fig:Covariant-Subtyping}.

\begin{figure}
\includegraphics[scale=0.6]{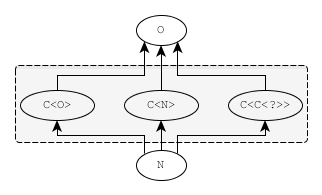}

\protect\caption{\label{fig:Invariant-Subtyping}Subtyping (between types of rank 0
and 1) due to invariant subtyping}
\end{figure}

In order to appreciate the intricacy of the generic subtyping relation
in Java, we should now---before things get more complex---notice the
self-similarity that is getting evident in the relation, where the
subtyping relation between types inside the dotted part of the diagram
is the same as the relation between types in the input relation (the
initial subtyping relation, in Figure~\ref{fig:Initial-Subtyping}).
As we demonstrate shortly, contravariant subtyping results in a ``flipped''
relation (\emph{i.e.}, an opposite ordering relation) and invariant
subtyping results in a ``flattened'' relation (\emph{i.e.}, a discrete
ordering relation). This observation will get reinforced, but will
also get somewhat more blurred, as the construction process proceeds
further. (If we have \emph{only} the contravariant subtyping rule,
the resulting subtyping relation will be as in Figure~\ref{fig:Contravariant-Subtyping},
and if we have \emph{only} the invariant subtyping rule, the resulting
subtyping relation will be as in Figure~\ref{fig:Invariant-Subtyping}.)

\begin{figure}
\noindent \begin{centering}
\includegraphics[scale=0.6]{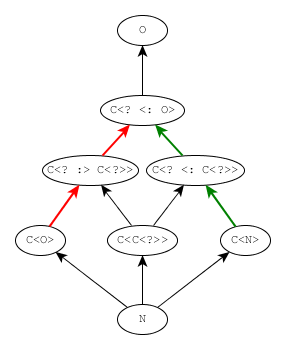}
\par\end{centering}

\protect\caption{\label{fig:Subtyping-rank-1}Subtyping between types of rank 0 and
1}
\end{figure}

To construct the full second version of the subtyping relation, resulting
from the first iteration of the construction of the relation \emph{with
all} \emph{three }subtyping rules, the construction process merges
the three earlier constructed relations by first identifying some
types in them (\emph{e.g.}, types \code{O} and \code{N}, and noting
that for generic types the type parameter `\code{?~:>~N}' is the
same as `\code{?~<:~O}', and `\code{?~:>~O}' is the same as
\code{O}') and consistently merging and defining all the subtype
relations between the identified types and other types in the subtyping
relations. Thus, with all three subtyping rules, the relation in Figure~\ref{fig:Subtyping-rank-1},
as the proper merging of the relations in Figure~\ref{fig:Covariant-Subtyping},
Figure~\ref{fig:Contravariant-Subtyping} and Figure~\ref{fig:Invariant-Subtyping},
presents the second, more accurate version of the full subtyping relation.

To construct the full and most accurate version of the subtyping relation
(\emph{i.e.}, the whole infinite relation), it should now be clear
that the process described is continued \emph{ad infinitum.} The purpose
of the Java subtyping operad we present in this paper is to formally
model each iteration in this construction process, based on the intuitions
we presented in the example above. We expand on this example in Section~\ref{sec:Applying-JSO},
after we present our subtyping operad in Section~\ref{sec:JSO}.

\subsection{Operads}

\begin{figure}
\noindent \begin{centering}
\includegraphics[angle=-90]{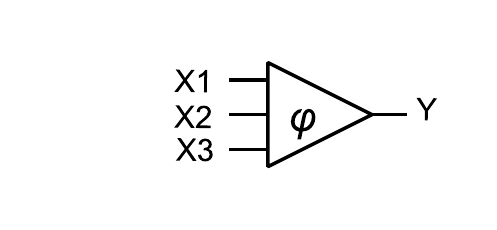}
\par\end{centering}

\protect\caption{\label{fig:An-Operad-Morphism}An Operad Morphism}
\end{figure}

Operads, from category theory, are used frequently to model self-similar
phenomena.\footnote{In some category theory literature the notion of operads we use is
called a `colored operad' or a `symmetric multicategory,' and
the notion is strongly related to the notion of a `monoidal category'.
In this paper we adopt the naming convention used by Spivak~\cite{spivak2014category}
and others, for reasons similar to theirs.} Informally, an operad is `a category whose morphisms are \emph{multiple-input}
single-output morphisms'. Thus, operads, as generalizations of categories,
embody in particular a generalization of the familiar notion of (single-input
single-output) morphisms. As such, a morphism of an operad is usually
depicted as in Figure~\ref{fig:An-Operad-Morphism}.

An operad, however, has some features that make it particularly suited
to model self-similarity~\cite{spivak2014category}, particularly
the composition formula for its generalized notion of morphisms. Similar
to regular categorical arrow/morphism composition, morphisms in operads
can be composed to define new morphisms. Composition of operad morphisms
is the source of most of the power of operads.

\begin{figure}
\noindent \begin{centering}
\includegraphics[scale=0.3]{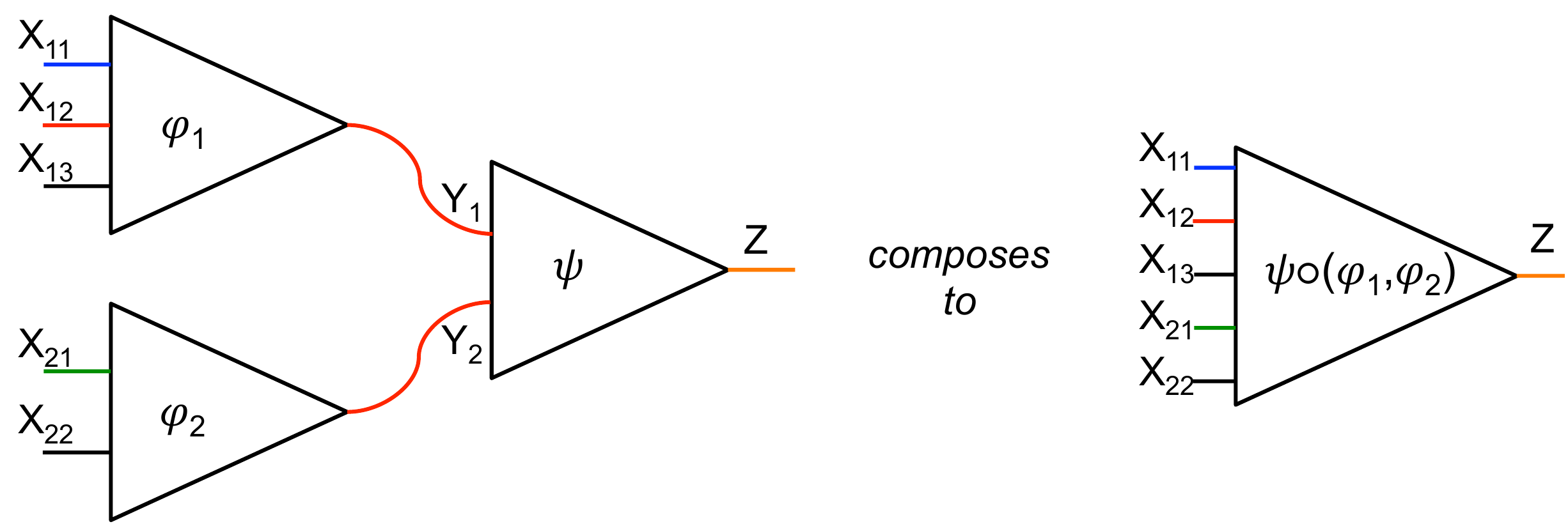}
\par\end{centering}

\protect\caption{\label{fig:Operad-Composition}Operad Composition ((c) 2014 David
Spivak)}
\end{figure}

The idea behind composition in operads is straightforward (see Figure~\ref{fig:Operad-Composition}).
The formal definition of operadic composition and of its associativity
requirement, however, are a bit more complicated than for a category,
since they involve much more variable indexing, so we do not present
these here, depending instead on the intuitiveness and the straightforwardness
of the idea behind them (the interested reader should consult~\cite[Sec. 7.4]{spivak2014category}).
More details about operads, including plenty of operad examples, can
be found in~\cite{spivak2014category} and~\cite{leinster2004higher}.

\section{\label{sec:JSO}$\protect\JSO$, A Simple Java Subtyping Operad}

The Java subtyping operad, which we call $\JSO$, is somewhat similar
to and has been inspired by the `wiring diagrams' operad Spivak
presents in~\cite{Spivak2013,spivak2014category}. The objects of
$\JSO$ are subtyping relations. Its morphisms include four relation
transformation morphisms named $copy$, $flip$, $flat$, and $merge$
(which we describe shortly) corresponding to the application of the
three generic subtyping rules and the merging step we described in
Section~\ref{sub:Subtyping-in-Java}, an \emph{identity }transformation
(as part of the requirements for defining an operad), and the compositions
of all composable (\emph{i.e.}, composition-compatible) combinations
of these transformations.

For each generic class \code{C}, in iteration $n$ of the construction
process the $copy$ morphism constructs types \code{C<?~<:~Type>}
(of rank $n+1$) for each type \code{Type} of the (rank $n$ and
lower) types in the input subtyping relation, and it defines the subtyping
relation between these types the same as between the corresponding
types in the input relation (hence the name $copy$). The $copy$
morphism thus models covariant subtyping.

Similarly, for each generic class \code{C}, the $flip$ morphism
constructs types \code{C<?~:>~Type>} for each input type, but reverses
the input subtyping relation between these (effecting an ordering
between the new types opposite to that between corresponding input
types; hence the name $flip$) to model contravariant subtyping. The
$flat$ morphism similarly constructs for each generic class \code{C}
the types \code{C<Type>} with no subtyping relation between these
new types (other than the trivial ordering due to reflexivity, effecting
a discrete ordering between the new types; hence the name $flatten$,
which we shorten to $flat$).

To produce their output subtyping relations, the three transformations
then, for each generic class \code{C}, ``embed'' the new subtyping
relation they defined between types constructed using \code{C} and
the input relation, where the relation gets embedded in place of type
\code{C<?>} in the initial subtyping relation (this embedding of
a relation inside another relation can be precisely defined using
operadic terms similar to ones in the `wiring diagrams' operad of
Spivak~\cite{Spivak2013,spivak2014category}).

\begin{figure}
\noindent \begin{centering}
\includegraphics[angle=-90,scale=0.7]{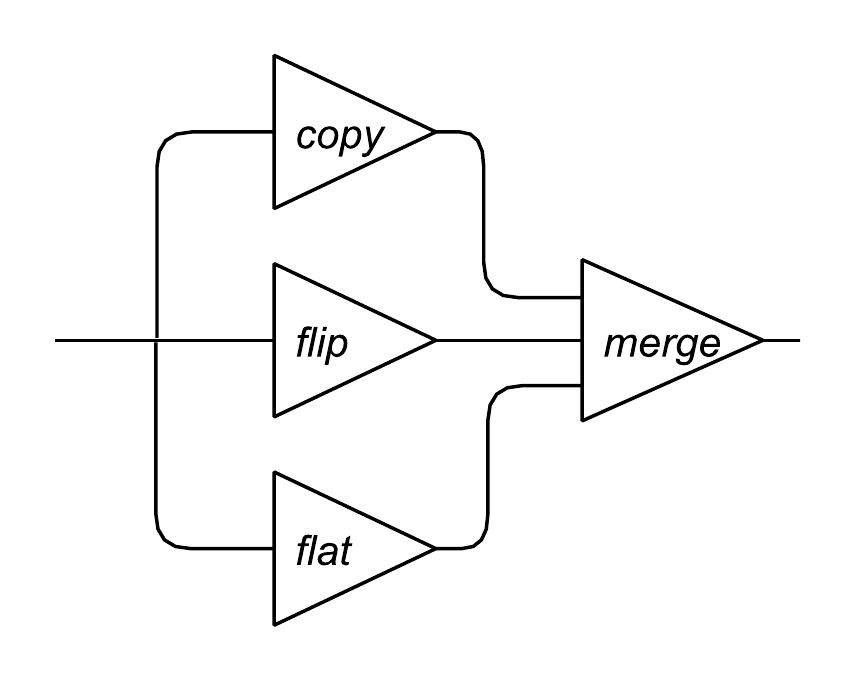}
\par\end{centering}

\protect\caption{\label{fig:JSO-Morphisms}The main morphisms of $\protect\JSO$, and
how they are composed to define successively more accurate versions
of the Java subtyping relation.}
\end{figure}

The $merge$ transformation is a little different than the three other
main transformations. The three subtyping relations that are input
to the $merge$ morphism will be the output relations of the three
morphisms $copy$, $flip$, and $flat$, as shown in Figure~\ref{fig:JSO-Morphisms},
\emph{i.e.}, $merge$ will be composed (operadically) with $copy$,
$flip$, and $flat$. The $merge$ morphism does not construct any
new types but rather identifies any repeated new types in its three
inputs while keeping their subtyping relations with other types (\emph{i.e.},
$merge$ performs a pushout/fibered coproduct of its input relations,
quotienting over identified types and relations between them, effecting
a ``union'' of its inputs). The $merge$ morphism thus merges the
three input subtyping relations into one output subtyping relation
(hence the name $merge$).

For example, on input the relation in Figure~\ref{fig:Initial-Subtyping},
the $copy$ morphism will have the relation in Figure~\ref{fig:Covariant-Subtyping}
as its output. Similarly, the morphisms $flip$ and $flat$ will have
the relations in Figure~\ref{fig:Contravariant-Subtyping} and Figure~\ref{fig:Invariant-Subtyping}
as their output, respectively. On input the three relations in Figures~\ref{fig:Covariant-Subtyping},
\ref{fig:Contravariant-Subtyping} and~\ref{fig:Invariant-Subtyping},
the $merge$ morphism will have the relation in Figure~\ref{fig:Subtyping-rank-1}
as its output. We further demonstrate $\JSO$ in Section~\ref{sec:Applying-JSO}.

Using the fact that subtyping relations can be viewed as DAGs (\emph{i.e.},
directed acyclic graphs), confirming that $\JSO$ (\emph{i.e.}, subtyping
relations and the announced transformations) is an operad is tedious
but relatively straightforward (similar to proving that directed graphs
and homomorphisms over them define a category, the category \textbf{Graph},
but involving more variable-indexing).

Having defined $\JSO$, each iteration of the construction process
of the Java subtyping relation (as described in Section~\ref{sub:Subtyping-in-Java})
can be concisely described by a single $\JSO$ morphism, which we
name $\JSM$. As such, we define
\[
\JSM:=merge\circ(copy,flip,flat)
\]
where $\circ$ is operadic composition (see~\cite[Sec. 7.4]{spivak2014category},
particularly Remark 7.4.1.2, for notation).

\paragraph{Bounded Type Variables}

In the description of $\JSO$ transformations we did not consider
type variable bounds, assuming instead that all type variables have
the default upper bound \code{O} (\emph{i.e.}, \code{Object}). Bounded
type variables can be handled in $\JSO$ using an additional \emph{clip}
morphism that takes the three output relations of $copy$, $flip$
and $flat$ and, for each generic class \code{C}, ``throws away''
(in an operadic way) the types (and their subtype relations) that
are not within the bounds of the type variable of \code{C} (remember
that, for simplicity, we are assuming a generic class has a single
type variable). So far, however, it is not clear to us how to handle
the case when the type variable ``appears in its own bound'' (\emph{i.e.},
the case when we have `F-bounded polymorphism'). Given the importance
of handling this case, we thus leave the modeling of bounds of type
variables altogether to future work.

\section{\label{sec:Applying-JSO}Examples of Applying $\protect\JSO$}

To demonstrate how $\JSO$ works, we present in this section two examples
of how $\JSO$ defines the subtyping relation for two simple Java
programs.

The first example furthers the construction of the subtyping relation
for the simple program we presented in Section~\ref{sub:Subtyping-in-Java}
to define the third version of the subtyping relation between types
of rank 0, 1 and 2 for that program. This version of the relation
will look as in Figure~\ref{fig:SubtypingC2}.

The second Java program example, in addition to the declaration of
class \code{C}, has a second generic class declaration

\code{class D<T> extends Object \{\}}.

The subclassing relation and the initial subtyping relation for this
program will look as in Figure~\ref{fig:SubtypingCD0}, while the
second version of the subtyping relation for this program will look
as in Figure~\ref{fig:SubtypingCD1}. See Appendix~\ref{sec:Demonstration-of-JSO}
for more details on these examples.

\section{\label{sec:Related-Work}Related Work}

The addition of generics to Java has motivated much research on generic
OOP and also on the type safety of Java and similar languages. Much
of this research was done before generics were added to Java\emph{.
}For example, the work in \cite{Bank96,Agesen97,Bracha98,Corky98,Thorup99}
was mostly focused on researching OO generics, while the work in \cite{Drossopoulou97,flatt98,drossopoulou99,flatt99}
was focused on type safety. Some research on generics, and generic
type inference, was also done after generics were added to Java\emph{,
e.g.},~\cite{SmithJava08,Zhang:2015:LFO:2737924.2738008,AbdelGawad2016c,Grigore2017}.

However, FJ/FGJ (Featherweight Java/Featherweight Generic Java)~\cite{FJ/FGJ}
is probably the most prominent work done on the type safety of Java,
including generics. FJ/FGJ did not consider variance annotations (or
wildcard type parameters) though.

Separately, probably as the most complex feature of Java generics,
the addition of ``wildcards'' (\emph{i.e.}, wildcard type parameters)
to Java (in~\cite{Torgersen2004}, which is based on the earlier
research in~\cite{Igarashi02onvariance-based}) also generated some
research that is particularly focused on modeling wildcards and variance
annotations~\cite{MadsTorgersen2005,Cameron2007,KennedyDecNomVar07,Cameron2008,Summers2010}.
This significant and substantial work points yet to the need for more
research on wildcards and generics.

The use of category theory tools in computer science is well-known~\cite{Pierce91,spivak2014category},
particularly in researching the semantics of programming languages~\cite{DTAL,Scott82,CatRecDomEqs82,GunterHandbook90,BauerCategory01,Gierz2003,DomTheoryIntro}.
After all, most of category theory has a constructive computational
``feel'' to it~\cite{Rydeheard1988,CatTheoryGentleIntro}. Some
category theory tools, particularly coalgebras, have also been used
to research OOP~\cite{CanningFbounded89,CookInheritance90,Jacobs95,Poll2000276}.
However, to the best of our knowledge, operads (or symmetric multicategories)~\cite{leinster2004higher}
have not been used before in programming languages research.\footnote{As we mentioned earlier, what we call operads are called `symmetric
multicategories' or `colored operads' in some category theory literature,
where an operad in this literature is a one-object multicategory.
In this paper we follow the naming convention of Spivak~\cite{spivak2014category}
and others.} Given the increasing awareness of its power and its wide array of
practical scientific applications, it is currently expected that the
use of category theory in computer science research (and other scientific
research) will further increase~\cite{lawvere2009conceptual,spivak2014category}.

\section{\label{sec:Conclusion-and-FW}Discussion and Future Work}

The simple operad $\JSO$ we presented in this paper seems to nicely
capture some of the main features of the generic subtyping relation
in Java and similar OO languages, particularly its self-similarity.
Based on our development of $\JSO$ as a model of generic Java subtyping
that particularly reveals its intricate self-similarity, we believe
that, generally speaking, using category theory (which has a rich
set of tools, including powerful notions such as operads) more may
hold the key to having a better understanding of complex features
of programming languages, such as wildcards and generics.

More specifically, we believe the self-similarity of the Java subtyping
relation (revealed by $\JSO$) is obscured by three factors: first,
the merging step (the work of the $merge$ morphism of $\JSO$) in
the construction of the subtyping relation, particularily its identification
of some of the types constructed by other components of $\JSO$, thereby
ambiguating the origin of these types. Second, we believe the self-similarity
is also obscured by the inexpressibilty of the \code{Null} type in
Java, which makes some of the types constructed by $\JSO$ involving
\code{Null} (\emph{e.g.}, type \code{C<N>} above), and their subtyping
relations, sound unfamiliar.

In our opinion, the third reason for obscuring the self-similarity
of the generic subtyping relation in Java is thinking about the relation
in \emph{structural-typing} terms rather than \emph{nominal-typing}
ones. Although it seems the polymorphic structural subtyping relation
(with variance annotations) exhibits self-similarity similar to the
one $\JSO$ demonstrates for Java, but it should be noted that nominal
typing in languages such as Java, C\# and Scala, particularly the
ensuing identification of type/contract inheritance with nominal subtyping~\cite{InhSubtyNWPT13,NOOPsumm,AbdelGawad2015},
are used in $\JSO$, first, to define the initial version of the nominal
subtyping relation in its iterative construction process (directly
based on the subclassing/inheritance relation), and, second, in the
embedding step (into the initial subtyping relation) in each iteration
of the process. A simple and strong connection between type inheritance
and subtyping does not exist when thinking about the Java subtyping
relation in structural typing terms. It seems to us that not making
this connection, keeping instead the subtyping relation separate and
independent from the inheritance relation, makes it harder to see
the self-similarity of generic nominal subtyping, its intricacies,
and its connections to the subclassing/inheritance relation.

Having said that, more work on $\JSO$ is needed however, to make
it model Java subtyping more accurately. Even though we presented
how $\JSO$ can model the three generic subtyping rules in Java, and
how they are combined to define the subtyping relation, $\JSO$ does
not model bounded type variables in particular. We hinted earlier
in this paper to how bounded type variables can be modeled, but the
actual definition of $\JSO$ to include them remains to be done.

Also, $\JSO$ can be made more general---covering more features of
the Java generic subtyping relation and also suggesting how generic
subtyping in Java can be extended---if $\JSO$ includes a notion of
\emph{type intervals}, which roughly are `intervals over the subtyping
relation'~\cite{AbdelGawad2014b}, thereby supporting types other
than \code{Object} and \code{Null} as upper and lower bounds, (1)
firstly, for type variables~\cite{SmithCompleting07,SmithJava08,SmithDesigning10},
and, (2) secondly, for generic type parameters (as such, type intervals
smoothly subsume and generalize wildcard types).

It may be useful also to extend $\JSO$ to model subtyping between
generic types that have type variables in them (in this paper we modeled
the subtyping relation between \emph{ground} generic types, which
have no type variables in them). Unifying the last two suggestions,
we believe it may be useful if $\JSO$ more generally supports a notion
of \emph{nominal intervals}~\cite{AbdelGawad2016c}, where type variables
are viewed as names for type intervals and where intervals with the
same lower and upper bounds but with different names are considered
unequal type intervals.

\bibliographystyle{plain}

\appendix

\section{\label{sec:Demonstration-of-JSO}Demonstrating the Java Subtyping
Operad}

In this appendix we present two examples demonstrating how $\JSO$,
as defined in this paper, defines the construction of the subtyping
relation in two sample Java programs with simple class declarations.

\subsection{Example 1: One Generic Class, and Types of Rank 0, 1 and 2}

\begin{figure*}
\noindent \begin{centering}
\includegraphics[scale=0.6]{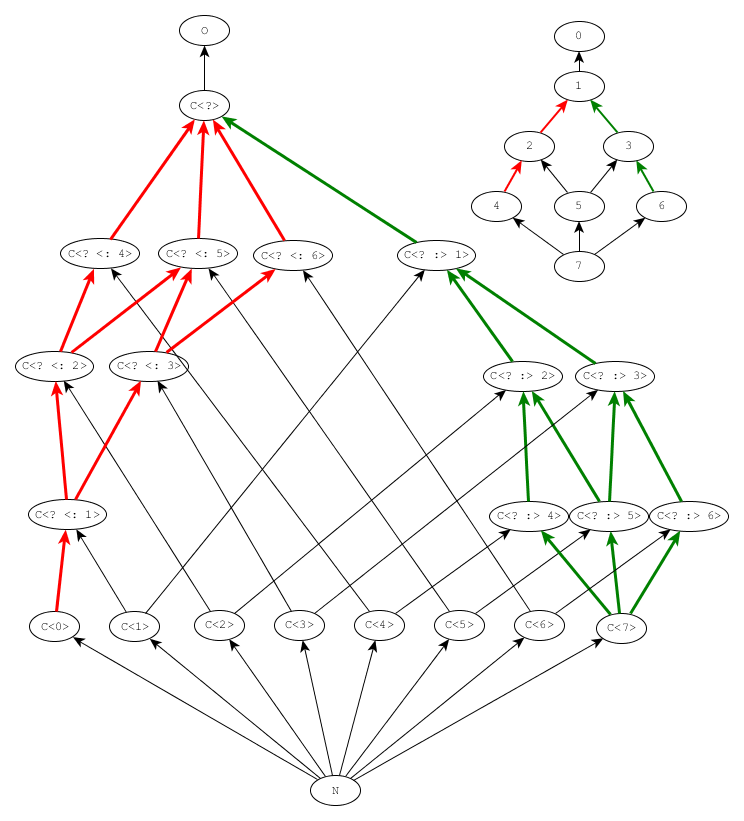}
\par\end{centering}

\protect\caption{\label{fig:SubtypingC2}Subtyping between rank 0, 1 and 2 types (with
one generic class)}
\end{figure*}

In this section we continue the example we started in Section~\ref{sub:Subtyping-in-Java}.
To shorten the type names, we use the numbers 0-7 as labels for the
types in Figure~\ref{fig:Subtyping-rank-1}, as presented in the
upper-right corner graph of Figure~\ref{fig:SubtypingC2}.

Using this labeling, the output subtyping relation resulting from
applying the $\JSM$ morphism of $\JSO$ to the relation in Figure~\ref{fig:Subtyping-rank-1}
as its input will be as depicted in the main graph of Figure~\ref{fig:SubtypingC2}.
The color highlighting in the graph helps see part of the effect of
the individual component morphisms of $\JSM$---$copy$ producing
the relations in green, $flip$ producing the relations in red, $flat$
producing the flat relation at the bottom (\emph{i.e.}, between the
``atom types'', right above type \code{N} in Figure~\ref{fig:SubtypingC2}),
and $merge$ producing the whole third version of the subtyping relation
for the sample Java program.

\subsection{Example 2: Two Generic Classes, and Types of Rank 0 and 1}

Assuming another Java program that has the same class declaration
for \code{C} but has a second generic class declaration

\begin{figure*}
\noindent \begin{centering}
\includegraphics[scale=0.6]{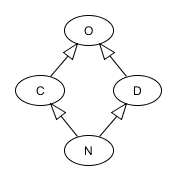}~~~\includegraphics[scale=0.6]{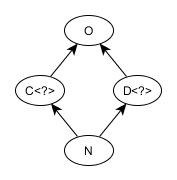}
\par\end{centering}

\protect\caption{\label{fig:SubtypingCD0}Subclassing and subtyping between rank 0
types (with two generic classes)}
\end{figure*}

\begin{figure*}
\noindent \begin{centering}
\includegraphics[scale=0.4]{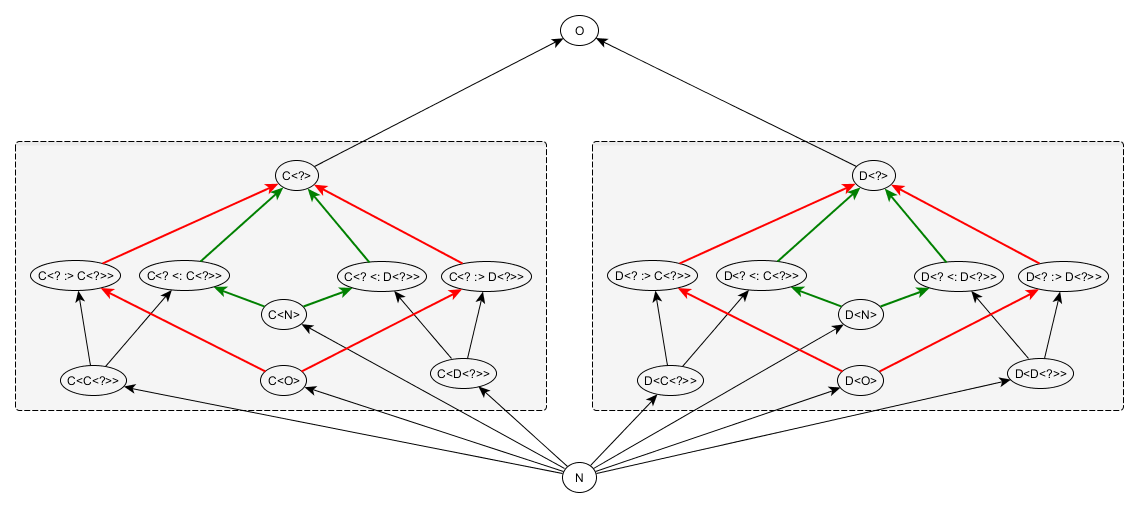}
\par\end{centering}

\protect\caption{\label{fig:SubtypingCD1}Subtyping between rank 0 and 1 types (with
two generic classes)}
\end{figure*}

\code{class D<T> extends Object \{\}}\\
we will then have the subclassing and initial subtyping relations
as in Figure~\ref{fig:SubtypingCD0}. Applying $\JSM$ to the relation
in Figure~\ref{fig:SubtypingCD0}, the resulting second version of
the relation will be as in Figure~\ref{fig:SubtypingCD1}.

The reader is encouraged to confirm that she sees the relations in
Figure~\ref{fig:SubtypingC2} and Figure~\ref{fig:SubtypingCD1}
as intuitive. She is also encouraged to construct the third version
of the subtyping relation for the second program, which results from
the application of $\JSM$ to the relation in Figure~\ref{fig:SubtypingCD1}.
\end{document}